\documentclass[reqno,11pt]{amsart}
\usepackage{amscd}
\usepackage{slashed}
\usepackage{amssymb}
\usepackage{ushort}
\usepackage{mathtools}
\usepackage{esint}
\usepackage{enumitem}
\usepackage[off]{auto-pst-pdf} % schnelle Version, Bilder müssen dazu schon vorliegen
\usepackage{pst-grad} % For gradients
\usepackage{pst-plot} % For axes
\usepackage[mathscr]{eucal}
\numberwithin{equation}{section}
\allowdisplaybreaks[1]

\title[Linear Stability of Rotating Black Holes]{Linear Stability of Rotating Black Holes: \\ Outline of the Proof}

\author[F.\ Finster]{Felix Finster}
\thanks{F.F.\ is supported in part by the Deutsche Forschungsgemeinschaft.}
\address{Fakult\"at f\"ur Mathematik \\ Universit\"at Regensburg \\ D-93040 Regensburg \\ Germany}
\email{Felix.Finster@mathematik.uni-regensburg.de}

\author[J.\ Smoller]{Joel Smoller \\ \\ September 2016}
\thanks{J.S.\ is supported in part by the National Science Foundation,
Grant No.\ DMS-1105189.}
\address{Mathematics Department \\ The University of Michigan \\ Ann Arbor, MI 48109, USA}
\email{smoller@umich.edu}

\newtheorem{Def}{Def.}[section]
\newtheorem{Thm}[Def]{Theorem}
\newtheorem{Prp}[Def]{Proposition}
\newtheorem{Lemma}[Def]{Lemma}

\newcommand{\Thanks}{\vspace*{.5em} \noindent \thanks}
\newcommand{\Proof}{\begin{proof}}
\newcommand{\QED}{\end{proof} \noindent}

\newcommand{\C}{\mathbb{C}}
\newcommand{\R}{\mathbb{R}}
\newcommand{\1}{\mbox{\rm 1 \hspace{-1.05 em} 1}}
\newcommand{\Z}{\mathbb{Z}}

\newcommand{\N}{\mathbb{N}}

\newcommand{\beq}{\begin{equation}}
\newcommand{\eeq}{\end{equation}}

\newcommand{\A}{\mathcal{A}}

\newcommand{\phiD}{\phi^\text{\tiny{\rm{D}}}}

\renewcommand{\H}{\mathscr{H}}

\newcommand{\Lin}{\text{\rm{L}}}

\newcommand{\D}{\mathscr{D}}
\newcommand{\scrM}{\mycal M}

\DeclareFontFamily{OT1}{rsfso}{}
\DeclareFontShape{OT1}{rsfso}{m}{n}{ <-7> rsfso5 <7-10> rsfso7 <10-> rsfso10}{}
\DeclareMathAlphabet{\mycal}{OT1}{rsfso}{m}{n}

\DeclareMathOperator{\im}{Im}

\begin{document}
\maketitle

\begin{abstract}
After a brief introduction to the black hole stability problem,
we outline our recent proof of the linear stability of the
non-extreme Kerr geometry.
\end{abstract}

\tableofcontents

\section{Introduction}
Black holes are peculiar, mysterious and probably the most exciting objects in our universe.
They arise mathematically from particular solutions of Einstein's equations of General Relativity (GR).
The first solutions describing black holes were found by Karl Schwarzschild in 1915
in his quest to describe stars by spherically symmetric and static solutions of the Einstein equations.
Shortly afterward, it was discovered that these solutions exhibit amazing physical properties,
the most interesting of which is that nothing, not even light, can escape from their interior.
In essence, this property defines a black hole.
Recently, astronomers have found strong evidence that black holes are ubiquitous in the
sense that they lie at the center of most galaxies.
These developments have brought black holes into the spotlight of astronomical research.
This has also inspired much interest in the study of theoretical properties of black holes.
In particular, the problem of stability of black holes is currently a question of high interest.
As Frolov and Novikov put it~\cite[page~143]{frolov+novikov},
this is ``one of the few truly outstanding problems that remain in the field of black
hole perturbations.''
We here report on recent work, partly carried out at the Center of Mathematical Sciences and
Applications at Harvard University, which shows that
rotating black holes are indeed stable under small perturbations (linear stability).

\section{The Kerr Black Hole}
In General Relativity, gravity is described geometrically in terms of
the curvature of space-time. Thus space-time is modelled mathematically
by a four-dimensional Lorentzian manifold~$(\scrM, g)$ of signature~$(+ \ \!\! - \ \!\! - \ \! - )$
(for a more detailed introduction see~\cite{bull} or the textbooks on GR~\cite{adler, misner, wald, straumann2}). In GR, Newton's gravitational law is replaced by the Einstein equations
\[ R_{jk} - \frac{1}{2}\: R\, g_{jk} = 8 \pi \kappa\, T_{jk} \:, \]
where~$R_{jk}$ is the Ricci tensor, $R$ is scalar curvature, and~$\kappa$ denotes the gravitational constant.
Here~$T_{jk}$ is the energy-momentum tensor which describes the distribution of matter in space-time.

A rotating black hole is described by the celebrated Kerr geometry. It is
a solution of the vacuum Einstein equations discovered in 1963 by Roy Kerr.
In the so-called Boyer-Lindquist coordinates, the Kerr metric is given by
(see~\cite{chandra, oneill})
\[ ds^{2} = \frac{\Delta}{U} \,( dt-a\sin^{2}\vartheta
\,d\varphi)^{2}-U \left( \frac{dr^{2}}{\Delta}+d\vartheta^{2}
\right) -\frac{\sin^{2}\vartheta}{U} \Big( a\,dt-(r^{2}+a^{2})d\varphi
\Big)^{2}, \]
where
\[ U = r^{2}+a^{2}\cos^{2}\vartheta,\qquad \Delta = r^{2}-2Mr+a^{2} \:, \]
and the coordinates~$(t,r, \vartheta, \varphi)$ are in the range
\[ -\infty<t<\infty,\;\;\;
M+{\sqrt{M^{2}-a^{2}}}<r<\infty,\;\;\; 0<\vartheta<\pi,\;\;\;
0<\varphi<2\pi \:. \]
Here the parameters~$M$ and~$aM$ describe the mass and the angular
momentum of the black hole. It is easily verified that in the
case~$a=0$, one recovers the Schwarzschild metric
\[ ds^{2} = \left(1-\frac{2M}{r} \right) dt^{2}-
\left(1-\frac{2M}{r} \right)^{-1} dr^2 -r^{2}(d\theta^{2}+\sin^{2}\theta\,d\varphi^{2})\:. \]

For the Kerr metric to describe a physical black hole we need to assume
that~$M^{2}> a^{2}$, giving a bound for the angular momentum
relative to the mass. In this so-called {\em{non-extreme case}},
the hypersurface
\[ r = r_{1} := M+{\sqrt{M^{2}-a^{2}}} \]
defines the {\em{event horizon}} of the black hole.
For~$r<r_1$, the radial coordinate~$r$ becomes time, whereas~$t$ becomes a spatial coordinate.
Since time always propagates to the future, the event horizon can be regarded as the ``boundary of no escape.''
Here we shall only consider the region~$r > r_1$ outside the event horizon.
The coefficients of the Kerr metric are independent of~$t$ and~$\varphi$,
showing that the space-time geometry is {\em{stationary}} and {\em{axi-symmetric}}.
One of the key features of the Kerr geometry is the existence of an {\em
ergosphere}, a region which lies outside the event horizon and
in which the vector field $\frac{\partial}{\partial t}$
is {\em space-like}.

It is a major open problem whether the Kerr geometry is stable under small perturbations.
This question is of physical relevance because if black holes were unstable, they would
decay and should therefore most probably not be observable in our universe.
A possible scenario is that a rotating black hole loses more and more angular momentum
by emitting gravitational radiation. If this were the case, the Kerr black hole with~$a>0$ would be unstable,
because asymptotically for large times it would settle down to the
static and spherically symmetric Schwarzschild solution with~$a=0$.

When analyzing the stability of the Kerr solution, one must specify which type of
perturbations are considered. Perturbing in the class of vacuum solutions corresponds
to perturbations by {\em{gravitational radiation}}. But one can also consider solutions
with matter, for example by considering the Einstein-Maxwell equations or the Einstein
equations coupled to other matter fields.
For the stability problem, one considers the Cauchy problem for initial data
which is close to a Kerr solution (in a suitable weighted Sobolev norm).
Then the question is whether the resulting solution settles down to a Kerr solution
asymptotically for large time.
Due to the nonlinearity of the Einstein equations, this problem is very difficult.
As a first step, one can linearize the equations around a Kerr solution and
analyze the long-time behavior of the resulting linear wave equations. This
is referred to as the {\em{linear stability problem}} for the Kerr black hole.

The problem of linear stability of the Schwarzschild black hole has a long history.
It goes back to the study by Regge-Wheeler~\cite{reggestab} who showed
that an integral norm of the perturbation of each angular mode is bounded uniformly in time.
Decay of these perturbations was first proved in~\cite{friedmanstab}.
For the Kerr black hole, linear stability has been an open problem for many years.
We now state our recent result on the linear stability of the Kerr black hole
and give an outline of the proof.

\section{The Teukolsky Equation and its Separation}
The problem of linear stability of black holes can be stated mathematically as the question of
whether solutions of massless linear wave equations in the Kerr geometry decay in time.
The different types of equations are characterized systematically in the Newman-Penrose
formalism by their spin~$s$, taking the possible values~$s=0, \frac{1}{2}, 1, \frac{3}{2}, 2, \ldots$.
From the physical point of view, the most interesting cases are $s=1$ (Maxwell field)
and~$s=2$ (gravitational waves). 
A general framework for analyzing the equations of arbitrary spin
in the Kerr geometry is due to Teukolsky~\cite{teukolsky}, who
showed that the massless equations of any spin can be rewritten as a single wave equation for
a complex scalar field~$\phi$. The Teukolsky equation reads
\begin{align*}
&\bigg( \frac{\partial}{\partial r} \Delta \frac{\partial}{\partial r} - \frac{1}{\Delta}
\left\{ (r^2+a^2)\: \frac{\partial}{\partial t} + a\: \frac{\partial}{\partial \varphi} - (r-M) \,s \right\}^2 
- 4 s \: (r+i a \cos \vartheta)\: \frac{\partial}{\partial t} \\
&\qquad + \frac{\partial}{\partial \cos \vartheta} \:\sin^2 \vartheta\: \frac{\partial}{\partial \cos \vartheta} 
+ \frac{1}{\sin^2 \vartheta} \left\{ a \sin^2 \vartheta\: \frac{\partial}{\partial t} 
+ \frac{\partial}{\partial \varphi} + i s \cos \vartheta \right\}^2 \bigg) \phi = 0 \:.
\end{align*}

We consider the Cauchy problem for the Teukolsky equation. Thus we seek a
solution~$\phi$ of the Teukolsky equation for given initial data
\[ \phi|_{t=0} = \phi_0 \qquad \text{and} \qquad \partial_t \phi|_{t=0} = \phi_1 \:. \]
Being a linear hyperbolic PDE, the Cauchy problem for the Teukolsky
equation has unique global solutions.
Also, taking smooth initial data, the solution is smooth for all time.
Our main interest is to show that solutions decay for large time.
In order to avoid specifying decay assumptions at the event horizon and at spatial infinity,
we restrict attention to compactly supported initial data outside the event horizon,
\beq \label{initsmooth}
\phi_0, \phi_1 \in C^\infty_0\big((r_1, \infty) \times S^2 \big) \:.
\eeq

Since the Kerr geometry is axisymmetric, the Teukolsky equation decouples into separate
equations for each azimuthal mode. Therefore, the solution of the Cauchy problem
is obtained by solving the Cauchy problem for each azimuthal mode and taking the sum of the resulting solutions.
With this in mind, we restrict attention to the Cauchy problem for a single azimuthal mode, i.e.
\beq \label{fouriermode}
\phi_0(r, \vartheta, \varphi) = e^{-i k \varphi}\: \phi_0^{(k)}(r, \vartheta)\:,\qquad
\phi_1(r, \vartheta, \varphi) = e^{-i k \varphi}\: \phi_1^{(k)}(r, \vartheta)
\eeq
for given~$k \in \Z/2$ (if~$s$ is half integer, then so is~$k$).
The main result of~\cite{stable} is stated as follows:
\begin{Thm} \label{thmdecay}
Consider a non-extreme Kerr black hole of mass~$M$ and angular momentum~$aM$
with~$M^2>a^2>0$. Then for any~$s \geq \frac{1}{2}$ and any~$k \in \Z/2$, the solution of the
Teukolsky equation with initial data of the form~\eqref{initsmooth} and~\eqref{fouriermode}
decays to zero in~$L^\infty_\text{\rm{loc}}((r_1, \infty) \times S^2)$.
\end{Thm} \noindent
This theorem establishes in the dynamical setting that the non-extreme Kerr black hole
is linearly stable.

Before outlining the proof of this theorem,
we remark that in the case~$a=0$
of the Schwarzschild geometry, the above result was already obtained in the paper~\cite{schdecay};
this was our starting point for attacking the problem with~$a>0$.

Restricting attention to a fixed azimuthal mode, the Teukolsky equation becomes
\begin{align*}
&\bigg( \frac{\partial}{\partial r} \Delta \frac{\partial}{\partial r} - \frac{1}{\Delta}
\left\{ (r^2+a^2)\: \frac{\partial}{\partial t} -i a k - (r-M) \,s \right\}^2 
- 4 s \: (r+i a \cos \vartheta)\: \frac{\partial}{\partial t} \\
&\qquad + \frac{\partial}{\partial \cos \vartheta} \:\sin^2 \vartheta\: \frac{\partial}{\partial \cos \vartheta} 
+ \frac{1}{\sin^2 \vartheta} \left\{ a \sin^2 \vartheta\: \frac{\partial}{\partial t} 
-ik + i s \cos \vartheta \right\}^2 \bigg) \phi = 0 \:.
\end{align*}
This equation can be further separated. Namely, making the separation ansatz
\[ \phi(t,r, \vartheta) = \frac{1}{\sqrt{r^2+a^2}}\: e^{-i \omega t} \,X(r)\, Y(\vartheta) \:, \]
the Teukolsky equation gives rise to the coupled system of ODEs
\beq \label{coupled}
{\mathcal{R}}_\omega X(r) = -\lambda X(r) \:,\qquad
\A_\omega Y(\vartheta) = \lambda Y(\vartheta) \:,
\eeq
where~$\lambda$ is a separation constant, and the radial operator~${\mathcal{R}}_\omega$ as well as
the angular operator~$\A_\omega$ are given by
\begin{align}
{\mathcal{R}}_\omega &= -\frac{(r^2+a^2)^2}{\Delta} \left(\frac{\partial^2}{\partial u^2}
- \frac{ \partial_u^2 \sqrt{r^2+a^2}}{\sqrt{r^2+a^2}} \right) \notag \\
&\qquad + \frac{1}{\Delta} \Big( -i \omega \,(r^2+a^2) - iak - (r-M) \,s \Big)^2 - 4 i s r \omega + 4 k\, a \omega \label{Rop} \\
\A_\omega &= - \frac{\partial}{\partial \cos \vartheta} \:\sin^2 \vartheta\: \frac{\partial}{\partial \cos \vartheta} 
+\frac{1}{\sin^2 \vartheta} \Big( -a \omega \sin^2 \vartheta +k - s \cos \vartheta \Big)^2 \:. \label{Aop}
\end{align}
Here the time separation reflects the fact that the Kerr metric is stationary.
However, as the Kerr metric is only axisymmetric (and not spherically symmetric),
the separation of the $\vartheta$-dependence does not correspond to
a symmetry of space-time. This separability goes back to the discovery by Carter
for the scalar wave equation~\cite{cartersep}.

We point out that, in contrast to the $\varphi$-dependence~\eqref{fouriermode},
the separation of the $\vartheta$-dependence cannot be performed for the initial data.
The reason is that the angular operator~$\A_\omega$ depends on~$\omega$.
Therefore, the decomposition into~$\vartheta$-modes makes it necessary to
restrict attention to fixed~$\omega$, meaning that the time dependence is that of a plane wave.
However, decomposing the wave into a superposition of plane waves~$e^{-i \omega t}$
would make it necessary to know the entire time evolution.

\section{Hamiltonian Formulation and Integral Representations}
In order to analyze the dynamics of the Teukolsky wave, it is useful to work with contour integrals
over the resolvent, as we now outline. In preparation, we write the Teukolsky equation in
Hamiltonian form. To this end, we introduce the two-component wave function
\[ \Psi = \sqrt{r^2+a^2} \begin{pmatrix} \phi \\ i \partial_t \phi \end{pmatrix} \]
and write the Teukolsky equation as
\beq \label{HPsi}
i \partial_t \Psi = H \Psi \:,
\eeq
where~$H$ is a second-order spatial differential operator.
We consider~$H$ as an operator on a Hilbert space~$\H$
with the domain
\[ \D(H) = C^\infty_0 \big( (r_1, \infty) \times S^2, \C^4 \big)\:. \]
We point out that the operator~$H$ is not symmetric on the Hilbert space~$\H$.
However, we choose the scalar product on~$\H$
as a suitable weighted Sobolev scalar product in such a way
that the operator~$H-H^*$ is bounded.

If~$H$ acted on a finite-dimensional vector space, the Cauchy problem for the
equation~\eqref{HPsi} with initial data~$\Psi_0$ could be solved with the Cauchy integral formula by
\beq \label{Cauchy}
\Psi(t) = -\frac{1}{2 \pi i} \ointctrclockwise_\Gamma e^{-i \omega t}\: \big(H-\omega)^{-1}\: \Psi_0\: d\omega\:,
\eeq
where~$\Gamma$ is a contour which encloses all eigenvalues of~$H$
(note that this formula holds for any matrix~$H$, even if it is not diagonalizable).
It turns out that in our infinite-dimensional setting, a similar formula holds.
Namely, using the fact that~$H-H^*$ is a bounded operator,
we prove that the resolvent~$R_\omega:= (H - \omega)^{-1}$
exists if~$\omega$ lies outside a strip enclosing the real axis (see~\cite[Lemma~4.1]{stable}):
\begin{Lemma} \label{lemmaresex}
For every~$\omega$ with
\[ |\im \omega| > c \:, \]
the resolvent~$R_\omega = (H-\omega)^{-1}$ exists and is bounded by
\[ \| R_\omega \| \leq \frac{1}{|\im \omega|-c}\:. \]
\end{Lemma} \noindent
When forming contour integrals, one must always make sure to stay outside
the strip~$|\im \omega| \leq c$, making it impossible to work with closed contours enclosing the spectrum.
However, we can work with unbounded contours in the following way (see~\cite[Corollary~5.3]{stable}):
\begin{Prp} For any integer~$p \geq 1$, the solution of the Cauchy problem for the Teukolsky equation
with initial data~$\Psi|_{t=0} = \Psi_0 \in \D(H)$ has the representation
\beq \label{propagatorp}
\Psi(t) = -\frac{1}{2 \pi i} \int_C e^{-i \omega t}\: \frac{1}{(\omega + 3 i c)^p} \;
\Big( R_\omega \,\big(H + 3 i c \big)^p \,\Psi_0  \Big)\: d\omega \:,
\eeq
where~$C$ is the contour
\beq \label{Cdef}
C = \big\{ \omega \:\big|\: \im \omega = 2c \big\} \cup 
\big\{ \omega \:\big|\: \im \omega = - 2c \big\}
\eeq
with counter-clockwise orientation.
\end{Prp} \noindent
Here the factor~$(\omega + 3 i c)^{-p}$ gives suitable decay for large~$|\omega|$
and ensures that the integral converges in the Hilbert space~$\H$.

\section{A Spectral Decomposition of the Angular Teukolsky Operator}
The next task is to use the separation of variables in the integrand of
our contour integral representation~\eqref{propagatorp}.
Regarding the angular equation~\eqref{coupled}
as an eigenvalue equation, we are led to consider
the angular operator~$\A_\omega$ in~\eqref{Aop} as an operator on the
Hilbert space
\[ \H_k := L^2(S^2) \cap \{ e^{-i k \varphi}\: \Theta(\vartheta) \:|\: \Theta : (0, \pi) \rightarrow \C \} \]
with dense domain~$\D(\A_\omega) = C^\infty(S^2) \cap \H_k$.
Unfortunately, the parameter~$\omega$ is not real but lies on the contour~\eqref{Cdef}.
As a consequence, the operator~$\A_\omega$ in~\eqref{Aop} is not symmetric,
because its adjoint is given by
\[ \A_\omega^* = \A_{\overline{\omega}} \neq \A_\omega \:. \]
The operator~$\A_\omega$ is not even a normal operator, 
making it impossible to apply the spectral theorem in Hilbert spaces.
Indeed, $\A_\omega$ does not need to be diagonalizable, because
there might be Jordan chains.
On the other hand, in order to make use of the separation of variables, we must decompose
the initial data into angular modes. This can be achieved by decomposing the
angular operator into invariant subspaces of bounded dimension, as is made
precise in the following theorem (see~\cite[Theorem~1.1]{tspectral}):
\begin{Thm} \label{thmmain}
Let~$U \subset \C$ be the strip
\[ |{\mbox{\rm{Im}}}\, \omega| < 3c \:. \]
Then there is a positive integer~$N$ and a family of bounded
linear operators~$Q^\omega_n$ on~$\H_k$
defined for all~$n \in \N \cup \{0\}$ and $\omega \in U$
with the following properties:
\begin{itemize}
\item[(i)] The image of the operator~$Q^\omega_0$ is an $N$-dimensional invariant
subspace of $\A_k$. 
\item[(ii)] For every $n\geq1$, the image of the operator~$Q^\omega_n$ is an at most
two-dimensional invariant subspace of $\A_k$.
\item[(iii)] The $Q^\omega_n$ are uniformly bounded in~$\Lin(\H_k)$,
i.e. for all $n \in \N \cup \{0\}$ and~$\omega \in U$,
\[ \|Q^\omega_n\| \leq c_2 \]
for a suitable constant $c_2=c_2(s,k,c)$ (here~$\| \cdot \|$ denotes the $\sup$-norm on~$\H_k$).
\item[(iv)] The~$Q^\omega_n$ are idempotent and mutually orthogonal in the sense that
\[ Q^\omega_n\, Q^\omega_{n'} = \delta_{n,n'}\: Q^\omega_n \qquad \text{for all~$n,n' \in \N \cup \{0\}$}\:. \]
\item[(v)] The $Q^\omega_n$ are complete in the sense that for every~$\omega \in U$,
\beq \label{strongcomplete}
\sum_{n=0}^\infty Q^\omega_n = \1
\eeq
with strong convergence of the series.
\end{itemize}
\end{Thm} \noindent

\section{Invariant Disk Estimates for the Complex Riccati Equation}
In order to locate the spectrum of~$\A_\omega$, we use detailed ODE estimates.
The operators~$Q^\omega_n$ are then obtained similar to~\eqref{Cauchy}
as Cauchy integrals,
\[ Q^\omega_n := -\frac{1}{2 \pi i} \ointctrclockwise_{\Gamma_n} s_\lambda\: d\lambda
\:,\qquad n \in \N_0 \:, \]
where the contour~$\Gamma_n$ encloses the corresponding spectral points,
and~$s_\lambda=(\A_\omega - \lambda)^{-1}$ is the resolvent of the angular operator.
What makes the analysis achievable is the fact that~$\A_\omega$ is an ordinary
differential operator. Transforming the angular equation in~\eqref{coupled} into
Sturm-Liouville form
\begin{equation} \label{5ode}
\left( -\frac{d^2}{du^2} + V(u) \right) \phi = 0 \:,
\end{equation}
(where~$u=\vartheta$ and~$V \in C^\infty((0, \pi), \C)$ is a complex potential),
the resolvent~$s_\lambda$ can be represented as an integral operator whose kernel
is given explicitly in terms of suitable fundamental solutions~$\phiD_L$ and~$\phiD_R$,
\beq \label{sldef}
s_\lambda(u,u') = \frac{1}{w(\phiD_L, \phiD_R)} \times \left\{
\begin{aligned} \phiD_L(u)\: \phiD_R(u') &\quad&& \text{if~$u \leq u'$} \\
\phiD_L(u')\: \phiD_R(u) &&& \text{if~$u' < u$}\:,
\end{aligned}  \right.
\eeq
where~$w(\phiD_L, \phiD_R)$ denotes the Wronskian.

The main task is to find good approximations for the solutions of the
Sturm-Liouville equation~\eqref{5ode} with rigorous error bounds
which must be uniform in the parameters~$\omega$ and~$\lambda$.
These approximations are obtained by ``glueing together'' suitable WKB, Airy and parabolic cylinder functions.
The needed properties of these special functions are derived in~\cite{special}.
In order to obtain error estimates, we combine several methods:
\begin{itemize}
\item[(a)] Osculating circle estimates (see~\cite[Section~6]{tspectral})
\item[(b)] The $T$-method (see~\cite[Section~3.2]{tinvariant})
\item[(c)] The $\kappa$-method (see~\cite[Section~3.3]{tinvariant})
\end{itemize}
The method~(a) is needed in order to separate the spectral points of~$\A_\omega$ 
(gap estimates).
The methods~(b) and~(c) are particular versions of {\em{invariant disk}} estimates
as derived for complex potentials in~\cite{invariant} (based on previous estimates for real potentials
in~\cite{angular} and~\cite{wdecay}).
These estimates are also needed for the analysis of the radial equation, see Section~\ref{secres} below.
We now explain the basic idea behind the invariant disk estimates.

Let~$\phi$ be a solution of the Sturm-Liouville equation~\eqref{5ode} with
a complex potential~$V$. Then the function~$y$ defined by
\[ y = \frac{\phi'}{\phi} \]
is a solution of the Riccati equation
\begin{equation} \label{riccati}
y'= V-y^2\:.
\end{equation}
Conversely, given a solution~$y$ of the Riccati equation,
a corresponding fundamental system for the Sturm-Liouville equation is obtained
by integration. With this in mind, it suffices to construct a particular approximate
solution~$\tilde{y}$ and to derive rigorous error estimates.
The invariant disk estimates are based on the observation that the Riccati
flow maps disks to disks (see~\cite[Sections~2 and~3]{invariant}).
In fact, denoting the center of the disk by~$m \in \C$ and its radius by~$R>0$, we get the
flow equations
\begin{align*}
R' &= -2 R \; {\mbox{\rm{Re}}}\, m \\
m' &= V - m^2 - R^2 \:.
\end{align*}
Clearly, this system of equations is as difficult to solve as the original Riccati equation~\eqref{riccati}.
But suppose that~$m$ is an approximate solution in the sense that
\begin{align*}
R' &= -2 R \; {\mbox{\rm{Re}}}\, m + \delta R \\
m' &= V - m^2 - R^2 \:+\: \delta m\:,
\end{align*}
with suitable error terms~$\delta m$ and~$\delta R$, then the Riccati flow
will remain inside the disk provided that its radius grows sufficiently fast, i.e.\
(see~\cite[Lemma~3.1]{invariant})
\[ \delta R \geq |\delta m|\:. \]
This is the starting point for the invariant disk method.
In order to reduce the number of free functions, it is useful to
solve the linear equations in the above system of ODEs by integration.
For more details we refer the reader to~\cite{invariant, tinvariant}.

\section{Separation of the Resolvent and Contour Deformations} \label{secres}
The next step is to use the spectral decomposition of the angular operator in Theorem~\ref{thmmain}
in the integral representation of the solution of the Cauchy problem.
More specifically, inserting~\eqref{strongcomplete} into~\eqref{propagatorp} gives
\beq \label{propagator2}
\Psi(t) = -\frac{1}{2 \pi i} \int_C \;\sum_{n=0}^\infty e^{-i \omega t}\: \frac{1}{(\omega + 3 i c)^p} \;
\Big( R_\omega \,Q^\omega_n \,\big(H + 3 i c \big)^p \,\Psi_0  \Big)\: d\omega \:.
\eeq
At this point, the operator product~$R_\omega Q^\omega_n$ can be expressed in
terms of solutions of the radial and angular ODEs~\eqref{coupled} which arise in
the separation of variables (see~\cite[Theorem~7.1]{stable}).
Namely, the operator~$Q^n_\omega$ maps onto an invariant subspace of~$\A_\omega$
of dimension at most~$N$, and it turns out that the operator product~$R_\omega \,Q^\omega_n$ leaves this
subspace invariant. Therefore, choosing a basis of this invariant subspace,
the PDE~$(H-\omega)R_\omega Q^n_\omega = Q^n_\omega$ can be rewritten as
a radial ODE involving matrices
of rank at most~$N$. The solution of this ODE can be expressed explicitly in terms
of the resolvent of the radial ODE. In order to compute this resolvent, it is useful to also transform the radial
ODE into Sturm-Liouville form~\eqref{5ode}.
To this end, we introduce the Regge-Wheeler coordinate~$u \in \R$ by
\[ \frac{du}{dr} = \frac{r^2+a^2}{\Delta} \:, \]
mapping the event horizon to $u=-\infty$. Then the radial ODE takes
again the form~\eqref{5ode}, but now with~$u$ defined on the whole real axis.
Thus the resolvent can be written as an integral operator with
kernel given in analogy to~\eqref{sldef} by
\[ s_\omega(u,v) = \frac{1}{w(\acute{\phi}, \grave{\phi})} \:\times\:
\left\{ \begin{array}{cl} \acute{\phi}(u)\, \grave{\phi}(v) & {\mbox{if~$v \geq u$}} \\[0.3em]
\grave{\phi}(u)\, \acute{\phi}(v) & {\mbox{if~$v < u$\:,}} \end{array} \right. \]
where~$\acute{\phi}$ and~$\grave{\phi}$ form a specific fundamental system
for the radial ODE.
The solutions~$\acute{\phi}$ and~$\grave{\phi}$ are constructed as
Jost solutions, using methods of one-dimensional scattering theory
(see~\cite{alfaro+regge} and~\cite[Section~6]{stable}, \cite[Section~3]{wdecay}).

The next step is to deform the contour in the integral representation~\eqref{propagator2}.
Standard arguments show that the integrand in~\eqref{propagator2} is holomorphic
on the resolvent set (i.e.\ for all~$\omega$ for which the resolvent~$R_\omega$ in~\eqref{propagatorp}
exists).
Thus the contour may be deformed as long as it does not cross singularities of the resolvent.
Therefore, it is crucial to show that the integrand in~\eqref{propagator2} is meromorphic
and to determine its pole structure.
Here we make essential use of Whiting's mode stability result~\cite{whiting}
which states, in our context, that every summand in~\eqref{propagator2} is holomorphic
off the real axis. In order to make use of this mode stability, we need to interchange the
integral in~\eqref{propagator2} with the infinite sum.
To this end, we derive estimates which show that the summands in~\eqref{propagator2}
decay for large in~$n$ uniformly in~$\omega$. Here we again use
ODE techniques, in the same spirit as described above for the angular equation
(see~\cite[Section~10]{stable}).
In this way, we can move the contour in the lower half plane arbitrarily close to the real axis. Moreover,
the contour in the upper half plane may be moved to infinity. We thus obtain
the integral representation (see~\cite[Corollary~10.4]{stable})
\[ \Psi(t) = -\frac{1}{2 \pi i} \sum_{n=0}^\infty\:\lim_{\varepsilon \searrow 0}
\int_{\R - i \varepsilon} \frac{e^{-i \omega t}}{(\omega + 3 i c)^p} \;
\Big( R_{\omega,n}\:Q_n^\omega \,\big(H + 3 i c \big)^p \,\Psi_0  \Big)\: d\omega \:. \]

The remaining issue is that the integrands in this representation might have 
poles on the real axis.
These so-called {\em{radiant modes}} are ruled out by a causality argument 
(see~\cite[Section~11]{stable}). We thus obtain the following result (see~\cite[Theorem~12.1]{stable}).
\begin{Thm} \label{thmrep}
For any~$k \in \Z/2$, there is a parameter~$p>0$ such that for any~$t<0$, the solution of the Cauchy problem 
for the Teukolsky equation with initial data
\[ \Psi|_{t=0} = e^{-i k \varphi}\: \Psi_0^{(k)}(r, \vartheta) \qquad \text{with} \qquad
\Psi^{(k)}_0 \in C^\infty(\R \times S^2, \C^2) \]
has the integral representation
\beq \begin{split}
\Psi&(t,u,\vartheta, \varphi) \\
&= -\frac{1}{2 \pi i} \:e^{-i k \varphi}\: \sum_{n=0}^\infty \int_{-\infty}^\infty \frac{e^{-i \omega t}}{(\omega + 3 i c)^p} \;
\Big( R^-_{\omega,n} \:Q_n^\omega
\big(H + 3 i c \big)^p \,\Psi_0^{(k)}  \Big)(u, \vartheta)\: d\omega \:,
\end{split} \label{propfinal}
\eeq
where~$R^-_{\omega,n} \Psi := \lim_{\varepsilon \searrow 0} \big(R_{\omega-i \varepsilon,n} \Psi)$.
Moreover, the integrals in~\eqref{propfinal} all exist in the Lebesgue sense.
Furthermore, for every~$\varepsilon>0$ and~$u_\infty \in \R$, there is~$N$ such that
for all~$u<u_\infty$,
\beq \label{biges}
\sum_{n=N}^\infty \int_{-\infty}^\infty \bigg\| \frac{1}{(\omega + 3 i c)^p} \;
\Big( R^-_{\omega,n} \:Q_n^\omega
\,\big( H + 3 i c )^p \,\Psi_0^{(k)} \Big)(u) \bigg\|_{L^2(S^2)} \: d\omega < \varepsilon \:.
\eeq
\end{Thm}

\section{Proof of Decay} \label{secradial}
Theorem~\ref{thmdecay} is a direct consequence of the integral representation~\eqref{propfinal}
in Theorem~\ref{thmrep}. Namely, combining the estimate~\eqref{biges} with Sobolev methods,
one can make the contributions for large~$n$ pointwise arbitrarily small.
On the other hand, for each of the angular modes~$n=0,\ldots, N-1$, 
the desired pointwise decay as~$t \rightarrow -\infty$ follows from the Riemann-Lebesgue lemma.
For details we refer to~\cite[Section~12]{stable}.

\section{Concluding Remarks} \label{secoutlook}
We first point out that the integral representation of Theorem~\ref{thmrep} is
a suitable starting point for a detailed analysis for the dynamics of the solutions of the Teukolsky equation.
In particular, one can study decay rates (similar as worked out for massive Dirac waves in~\cite{wdecay})
and derive uniform energy estimates outside the ergosphere (similar as for scalar waves in~\cite{sobolev}).
Moreover, using the methods in~\cite{penrose}, one could analyze superradiance phenomena
for wave packets in the time-dependent setting.

Clearly, the next challenge is to prove {\em{nonlinear stability}} of the Kerr geometry.
This will make it necessary to refine our results on the linear problem, for example
by deriving weighted Sobolev estimates and by analyzing the $k$-dependence of our estimates.
Moreover, it might be useful to combine our methods and results with microlocal techniques.

\Thanks {{\em{Acknowledgments:}}
We would like to thank Shing-Tung Yau for his encouragement and support.
F.F.\ is grateful to the Center of Mathematical Sciences and Applications at
Harvard University for hospitality and support during his two visits from
September 15 until October 15, 2015, and again from March 15 until April 15, 2016.

%\bibliographystyle{amsplain}
%\bibliography{../../aarbeit/felix}
\providecommand{\bysame}{\leavevmode\hbox to3em{\hrulefill}\thinspace}
\providecommand{\MR}{\relax\ifhmode\unskip\space\fi MR }
\providecommand{\MRhref}[2]{%
  \href{http://www.ams.org/mathscinet-getitem?mr=#1}{#2}
}
\providecommand{\href}[2]{#2}

\end{document}